\documentclass[prb,twocolumn,superscriptaddress,fleqn]{revtex4-2}

\usepackage{amsmath,amssymb}
\usepackage{graphicx}
\usepackage{epstopdf}
\usepackage{bm}
\usepackage{xcolor}

\newcommand{\LAS}     {LaAgSb$_2$}
\newcommand{\la}	{$^{139}$La}

\newcommand{\kk} 	{$\mathcal{K}$}

\newcommand{\TCa} 	{$T_\text{CDW1}$}
\newcommand{\TCb} 	{$T_\text{CDW2}$}

\begin{document}

\title[]{NMR evidence for a Peierls transition in the layered square-net compound LaAgSb$_2$}

\author{Seung-Ho Baek}
\affiliation{Department of Physics, Changwon National University, Changwon 51139, Korea}
\affiliation{Department of Materials Convergence and System Engineering, Changwon National University, Changwon 51139, Korea}
\author{Sergey L. Bud'ko}
\affiliation{Ames Laboratory, U.S. Department of Energy and Department of Physics and Astronomy, Iowa State University, Ames, Iowa 50011, USA}
\author{Paul C. Canfield}
\affiliation{Ames Laboratory, U.S. Department of Energy and Department of Physics and Astronomy, Iowa State University, Ames, Iowa 50011, USA}
\author{F. Borsa}
\affiliation{Ames Laboratory, U.S. Department of Energy and Department of Physics and Astronomy, Iowa State University, Ames, Iowa 50011, USA}
\author{Byoung Jin Suh}
\email[]{bjsuh@catholic.ac.kr}
\affiliation{Department of Physics,The Catholic University of Korea, Bucheon 14662, Korea}

\date{\today}


\begin{abstract}

We measured the central ($1/2\leftrightarrow -1/2$) and first satellite ($\pm3/2\leftrightarrow \pm1/2$) lines of the \la\ NMR spectra as a function of temperature in \LAS, in order to elucidate the origin and nature of the charge-density-wave (CDW) transitions at $T_\text{CDW1}=207$ K and $T_\text{CDW2}=186$ K. In the normal state, the Knight shift \kk\ reveals a fairly linear relationship with decreasing temperature, which is ascribed to a pseudogap in the spin excitation spectrum, pointing towards the material being an unconventional metal. Upon further cooling, \kk\ decreases more steeply below \TCa, indicative of the partial Fermi surface gap opening on top of the pseudogap. 
The most remarkable finding in our study is a clear splitting of the satellite lines at \TCa\ observed for $H\parallel c$, whose temperature dependence behaves as the BCS order parameter in the weak-coupling limit, evidencing that the CDW transition induces the periodic lattice distortion. Our NMR findings therefore demonstrate that the CDW transition in \LAS\ is of Peierls type, being driven by the electronic instability in the vicinity of the Fermi level. 

\end{abstract}


\maketitle


\section{Introduction}

The charge-density-wave (CDW) phenomenon has been observed in numerous two-dimensional (2D) layered metallic compounds such as transition-metal dichalcogenides (TMDs) and high-$T_c$ cuprates, as well as quasi-one-dimensional (1D) metals \cite{wilson75, salvo79, gruener94, rossnagel11,chen16}. However, whereas CDW in 1D is well understood by the conventional Peierls' model \cite{peierls55} in which the Fermi surface nesting (FSN) drives a periodic lattice distortion (PLD), the origin of CDW in higher dimensions is much more complicated than in 1D \cite{johannes08a} and strongly material dependent \cite{zhu17}. 
As examples, rather than FSN, strong electron correlations appear to be a crucial ingredient for charge ordering in high-$T_c$ cuprates \cite{chen16}, or local electron-phonon coupling (EPC) may be a main driving force for the CDW in TMDs \cite{johannes06,calandra09}.

\LAS\ is a unique member of the layered Sb square-net family $R$AgSb$_2$ ($R$=rare earth) in space group $P4/nmm$ \cite{myers99a,myers99}, isostructural to the Fe-oxypnictide LaFeAsO \cite{kamihara08}. 
The square-net is featured by a shorter in-plane atomic distance ($a/\sqrt{2}$) than the lattice constant $a$, resulting in a twice larger unit cell in real space [see Fig.\,\ref{structure}(a)] or a folding of the Brillouin zone in reciprocal space \cite{klemenz19}. The resultant linear band crossings may occur near the Fermi level, allowing to host Dirac fermions \cite{young15}, as indeed observed in the Bi square-net, SrMnBi$_2$ \cite{park11}. 
Interestingly, unlike other members of $R$AgSb$_2$, most of which have a magnetically ordered ground state, \LAS\ is nonmagnetic and undergoes two successive CDW transitions (CDW1 at 207 K and CDW2 at 186 K) \cite{song03} while exhibiting unusual large linear magnetoresistance (MR) \cite{myers99a}, which may be attributed to the quantum limit of the Dirac fermions \cite{wang12e}.  
\LAS\ reveals not only a Dirac-cone-like band structure near the Fermi level \cite{wang12e,shi16}, but also well-nested segments of the Fermi surface directly linked to the Dirac cone, suggesting a close relationship between the Dirac fermions and the CDW ordering \cite{shi16}. 

In this work we have performed \la\ NMR to investigate the microscopic origin of the CDW transitions in \LAS. As \la\ has a large quadrupole moment, it could probe changes of both the local spin susceptibility and charge environments, via the Knight shift \kk\ and the quadrupole frequency $\nu_Q$. Our NMR data indicate that the CDW1 transition opens up a Fermi surface gap and causes the periodic lattice distortion, whose order parameter is well described by BCS mean-field theory, at the same time. Together with the observation of phonon softening at the CDW wave vector \cite{chen17, bosak21}, we conclude that \LAS\ is a rare quasi-2D material that undergoes a nesting-driven weak-coupling CDW transition.

\section{Experimental details}

High-quality \LAS\ single crystals were flux grown from a typical Sb rich self-flux, La$_{0.045}$Ag$_{0.091}$Sb$_{0.864}$, as described in Ref.\,\cite{myers99a}. The size of the crystal used in this NMR measurement is roughly $1\times 2\times 4$ mm$^3$. 

\la\ (nuclear spin $I=7/2$) NMR measurements were carried out at the external field of 4.7 T in the range of temperature 100--300 K. 
The signal intensity turned out very weak in our experiments, which is ascribed largely to the low filling factor owing to the plate-like single crystal within the NMR coil, and partly to small rf penetration in a metal that is limited by the skin depth, $\delta =\sqrt{2\rho/\omega\mu}$ where $\rho$ is the resistivity, $\omega$ is the angular frequency of the rf, and $\mu$ is the magnetic permeability \cite{bennet}. Due to the weak signal intensity, we focused solely on measuring the temperature dependence of the NMR spectra, without attempting to measure the spin-lattice relaxation rate, $T_1^{-1}$. Further, among the three pairs of satellites expected for $I=7/2$, we have measured the first satellite pair only, because it yields sufficient information regarding the electric field gradient and its temperature evolution at the \la\ in an axial symmetry. 
The \la\ NMR spectra were obtained by a conventional Hahn spin-echo technique with a typical $\pi/2$ pulse length of 3 $\mu$s.

\section{Results}

\subsection{Quadrupole-perturbed \la\ NMR spectra}

For a nuclear spin $I> 1/2$, there are central ($\frac{1}{2}\leftrightarrow -\frac{1}{2}$) and satellite transitions between the $m$th and $(m-1)$th levels ($m=-I,-I+1,\cdots, +I$). In an axial symmetric surrounding (asymmetry parameter $\eta=0$), the allowed transitions to first order are given by \cite{bennet}
\begin{equation}
	\begin{split}
\nu(m\leftrightarrow m-1)  =\; &\nu_0 (1+\mathcal{K}) \\
&+\frac{1}{2}\nu_Q(3\cos^2\theta -1)\left(m-\frac{1}{2}\right),
	\end{split}
	\label{quad1}
\end{equation}
where $\mathcal{K}$ is the Knight shift, $\nu_0$ is the unshifted Larmor frequency, $\nu_Q$ is the nuclear quadrupole frequency, and $\theta$ is the angle between the principal axis $z$ of the electric field gradient (EFG) and an external field $\mathbf{H}$. 

\begin{figure}
\centering
\includegraphics[width=\linewidth]{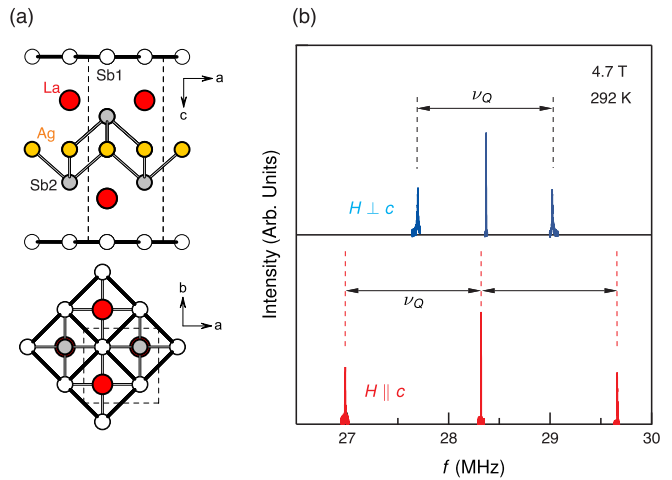}
\caption{ (a) Crystal structure of \LAS\ viewed along the $b$ (upper) and $c$ (lower) axes. The La atoms above and below the Ag-Sb2 plane are staggered, and the Ag and Sb1 atoms form an identical square-net, resulting in the $P4/nmm$ structure. The dashed lines denote the unit cell. (b) \la\ central and first satellite transitions for $H\perp c$ (upper) and $H\parallel c$ (lower) measured at 292 K and an external field of 4.7 T. The separations of the satellites for both directions are well described in terms of the single quadrupole frequency $\nu_Q=1.339$ MHz, as expected for an axial symmetry about the $c$ axis. 
}
\label{structure}
\end{figure}

Figure \ref{structure}(b) shows the central ($1/2\leftrightarrow -1/2$) and first satellite ($\pm3/2\leftrightarrow \pm1/2$) lines of the \la\ NMR spectra measured at $H=4.7$ T for $H\perp c$ and $H\parallel c$, respectively, at 292 K. 
The observed NMR spectra are excellently described by Eq.\,(\ref{quad1}), in which the distance between the first satellite lines is given by $\nu_Q$ and $2\nu_Q$ for $H\perp c$ and $H\parallel c$, respectively, indicating that the local symmetry at \la\ is indeed axial with respect to the principal axis of the EFG which lies along the $c$ axis. 
Interestingly, the line shape of each satellites appears to be asymmetric while the satellite pairs are symmetric about the central transition, similarly for both field orientations (see Fig.\,2). 
It should be noted that the FWHM of the NMR lines are very narrow, less than 10 kHz for the satellites which corresponds to only 1\% of $\nu_Q$ (see Fig.\,3), evidencing the very high homogeneity of our sample. Therefore, the asymmetric line shape of the satellites seems to reflect a feature of the satellite transitions ($\pm1/2\leftrightarrow \pm3/2$).

\subsection{Knight shift and Fermi surface gap}

\begin{figure*}
\centering
\includegraphics[width=\linewidth]{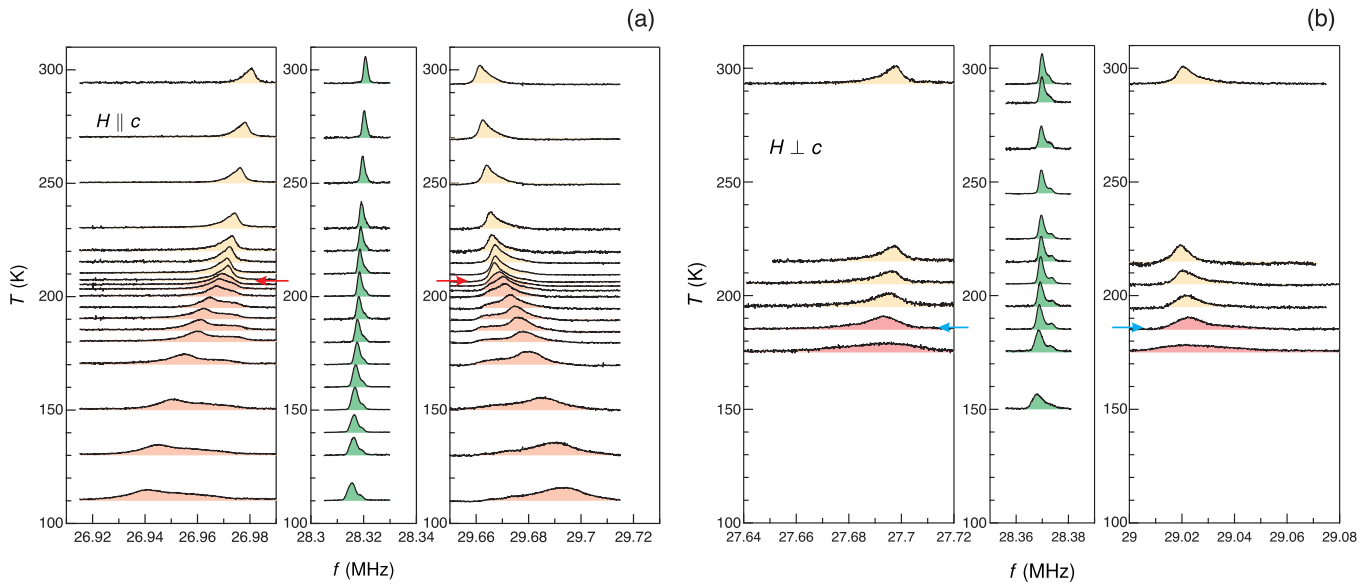}
\caption{Temperature dependence of \la\ NMR spectra for the central and first satellite transitions measured at $H=4.7$ T applied (a) parallel and (b) perpendicular to the $c$ axis. The left axis refers to the temperature at which each spectrum was obtained, whereas the amplitude of the spectra is plotted in arbitrary units. The central lines for both field directions similarly shift to lower frequency, i.e., the Knight shift is decreased,  with decreasing temperature. Whereas the satellite lines for $H\parallel c$ abruptly split below $T_\text{CDW1}=207$ K (marked by the red arrows), those for $H\perp c$ significantly broaden below $T_\text{CDW2}=186$ K (marked by the cyan arrows). 
}
\label{spec}
\end{figure*}

The temperature dependence of the NMR lines is presented in Fig.\,\ref{spec}. 
To begin with, we discuss the temperature dependence of the \la\ central line, shown in the middle panels of Figs.\,\ref{spec}(a) and \ref{spec}(b) for $H\parallel c$ and $H\perp c$, respectively. While the FWHM of the central line remains very narrow down to 100 K, the line clearly shifts to lower frequency with decreasing temperature, involving a weak line broadening. One can note that a small shoulder peak, which persists up to room temperature, is detected at the high-frequency side for both field directions. The small peak suggests that a small portion of the \la\ sites experiences slightly different hyperfine fields from the majority. Regardless of its origin and nature, 
the peak is so small that it could hardly affect the satellites, and it will be ignored for our purpose.

The Knight shift $\mathcal{K}\equiv (\nu-\nu_0)/\nu_0\times 100$ \% as a function of temperature is plotted in Fig.\,\ref{cdw}(a). For $H\perp c$ ($\theta=90^\circ$), the second order quadrupole shift of the central line given by $15\nu_Q^2/16\nu_0(1-\cos^2\theta)(1-9\cos^2\theta)$ for $I=7/2$ \cite{bennet}, was subtracted from the total shift.
In the normal state, we find that \kk\ decreases in an almost linear fashion with temperature, albeit very small, for both field orientations. 
For a simple Pauli metal, \kk\ or the local spin susceptibility is proportional to the density of states (DOS) at the Fermi level, $n(\epsilon_F)$, so it would have been expected to be independent of temperature in the normal state. Therefore, the unusual decrease of \kk\ may be ascribed to a pseudogap, which is frequently observed in many layered strange metals such as high-$T_c$ cuprates \cite{norman05}, Fe-based superconductors \cite{xu11,moon12}, and TMDs \cite{klemm00,borisenko08}. In fact, \LAS\ shows a rather atypical linear temperature dependence of resistivity in the normal state \cite{myers99a,song03}, and is believed to be a topological semimetal that hosts Dirac fermions \cite{wang12e,shi16}, being likely responsible for the pseudogap-like behavior. 

Upon further lowering temperature, we find that \kk\ decreases more rapidly just below \TCa, deviating from the linear temperature dependence in the normal state.  The small but clear additional reduction of \kk\ at \TCa\ observed for the two field directions suggests that $n(\epsilon_F)$ is further suppressed by the CDW1 transition,  greatly supporting that the CDW1 transition opens up a gap at the Fermi surface \cite{chen17}.
Interestingly, no additional notable change of \kk\ was observed at \TCb. This may suggest either that the CDW2 transition is too weak to yield a sizable gap, or that the nature of the CDW2 is different from that of the CDW1, as conjectured by the very different CDW wave vectors \cite{song03}. 

\subsection{Periodic lattice distortion in the CDW state}

Now we evaluate how the charge environment surrounding the \la\ nuclei evolves through the CDW transitions, via analysis on the temperature dependence of the satellites, giving rise to the information of the EFG and its spatial distribution. 

\begin{figure*}
\centering
\includegraphics[width=0.75\linewidth]{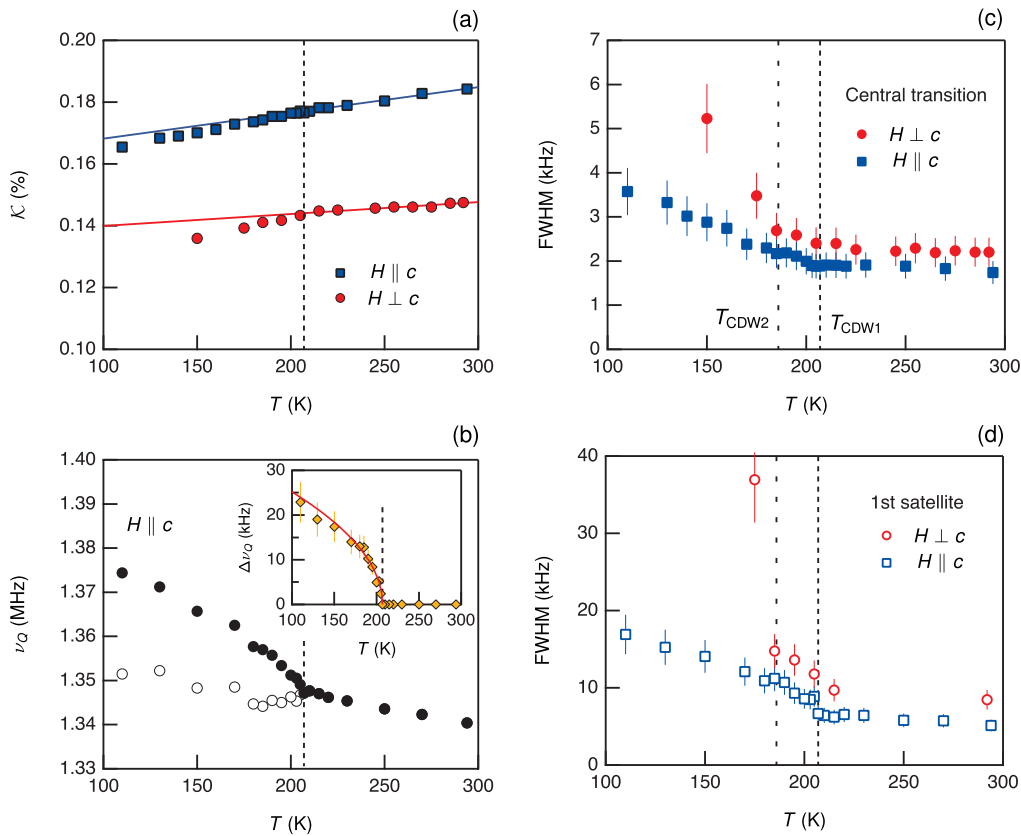}
\caption{(a) Temperature dependence of the Knight shift \kk. For both field directions, \kk\ decreases linearly with lowering temperature. Below \TCa\ the slope becomes larger suggesting a partial gap opening at the Fermi surface. (b) The quadrupole frequency $\nu_Q$ as a function of temperature. $\nu_Q$ sharply splits into two at \TCa. Inset shows an order parameter-like behavior of $\Delta\nu_Q$. The solid line is given by the weak coupling BCS theory. (c) and (d) FWHM for the central and first satellite lines as a function of temperature. Whereas the FWHM of all the NMR lines increases weakly for both field directions at \TCa,  a significant line broadening was observed below \TCb\ only for $H\perp c$.}
\label{cdw}
\end{figure*}

Most remarkably, we observed a clear splitting of the satellites for $H\parallel c$ just below \TCa, as shown in Fig.\,\ref{spec}(a). 
The $\nu_Q$ values extracted from the distance between the central and lower satellite lines are plotted in Fig.\,\ref{cdw}(b), which reveals the sharp splitting of the satellite lines. This means that the CDW1 transition induces two inequivalent \la\ sites which experience different EFG on average. 
Moreover, we confirmed that the difference of the $\nu_Q$ values between the split satellites, $\Delta\nu_Q$, serves as an order parameter for the CDW1 phase, which is consistent with the BCS mean field theory in the weak coupling limit, i.e., $\Delta\nu_Q \propto (1-T/T_\text{CDW1})^{1/2}$ [see the inset of Fig.\,\ref{cdw}(b)]. 
This result bears strong resemblance to the behavior of the CDW modulation intensity below \TCa\ observed by x-ray scattering measurements \cite{song03}, indicating that $\Delta\nu_Q$ is directly related to the lattice modulation associated with CDW order. That is, the local periodic lattice distortion emerges as a direct consequence of the charge modulation. 
Nevertheless, the presence of the two distinct \la\ sites is surprising, because the extremely long wavelength of the CDW modulation ($\lambda_1\sim 39a$) would result in a spatial distribution of the EFG at the \la\ sites, leading to an inhomogeneous quadrupole broadening rather than the split lines. 
This suggests that the PLD/CDW is formed in a way that discriminates the staggered La atoms above and below Ag-Sb2 plane in the unit cell [see Fig.\,\ref{structure}(a)]. 
%

In contrast, the satellites for $H\perp c$ are nearly unchanged through \TCa. The anisotropic changes of the spectra in the CDW1 phase may be understood by the strong dependence of the quadrupole effect on the angle between the direction of the EFG and $H$. 
 Since the lattice modulation for the CDW1 is developed along the $a$ axis \cite{song03}, $H$ is perpendicular to the EFG modulation which generates the two \la\ sites. In other words, $H$ has the same effects on the two \la\ sites, revealing the difference between them. However, the situation is quite different for $H\perp c$. Namely, the otherwise differentiated \la\ sites may feel a similar quadrupole-perturbed Zeeman field for $H\perp c$ because $H$ is applied along an arbitrary direction in the $ab$ plane in our experimental setup. 
 Note that the similar weak broadening observed for all the NMR lines below \TCa, as shown in Figs.\,\ref{cdw}(c) and \ref{cdw}(d), is what would be expected from a spatial distribution of the EFG. 

Another prominent feature found in Fig.\,2 is a significant line broadening of the NMR lines below the CDW2 transition temperature \TCb\ that occurs only for $H\perp c$ [see Figs.\,\ref{cdw}(c) and \ref{cdw}(d)]. Further, we find that the satellite lines for $H\perp c$ rapidly spread out in the CDW2 phase, becoming vanishingly small below $\sim 170$ K. 
While the external magnetic field anisotropy is naturally ascribed to the EFG modulation induced by the CDW2 modulation that is along the $c$ axis \cite{song03}, the considerable broadening of the satellite lines for $H\perp c$ suggests that the EFG distribution is much wider for the CDW2, probably due to the much shorter modulation wavelength of the CDW2 ($\lambda_2\sim 6.3c$) than the CDW1 ($\lambda_1\sim 39a$) \cite{song03}.


\section{Discussion}

We have established that the CDW transition at 207 K causes a (partial) gap opening at the Fermi level and generates two inequivalent \la\ sites, the difference in quadrupole frequency $\Delta\nu_Q$ of which is well described by the BCS mean field theory. 
%
These NMR findings microscopically prove that the CDW1 transition in \LAS\ is governed by the conventional (Peierls type) weak-coupling mechanism in which the Fermi surface nesting instability drives the periodic lattice distortion as well as the Fermi surface gap opening via electron-phonon interactions. This is consistent with the observation of acoustic phonon softening and the corresponding Kohn anomalies at the CDW wave vector \cite{chen17, bosak21}. 
%

It is worthwhile to compare the aforementioned with the recent NMR results obtained in the transition metal dichalcogenide, $2H$-TaSe$_2$, within which a strong-coupling mechanism driven by local electron-phonon coupling appears to underlie the CDW transition \cite{baek22}. The seemingly different CDW mechanism in the two layered 2D materials may be related to the structural motif of the TMDs and the square-net materials. Namely, unlike TMDs whose structural symmetry is lower than tetragonal, the square-net has an inherent instability against a Peierls transition, similar to 1D metals which are always unstable toward a lattice distortion \cite{peierls55}. Nevertheless, the square-net could be stabilized, e.g., by the overlap of $d$-orbitals from the transition-metal elements with the $s$ and $p$ orbitals in the square-net \cite{klemenz19}.

As to the CDW2 transition at 186 K, the anomalous anisotropic line broadening of the \la\ NMR spectra below 186 K reveals that the EFG modulation is formed along the $c$ axis due to the CDW2 ordering. However, neither a Fermi surface gap opening, although it is suggested by the resistivity anomaly for $H\parallel c$ \cite{song03}, nor the order parameter associated with the CDW2 was observed in our work, thus leaving the origin and nature of the CDW2 still an open question.

\section{Summary}

We carried out the \la\ NMR investigation in the Sb square-net \LAS. In the normal state, we found an unusual pseudogap behavior, reflecting the unconventional metallic state of the material. At the CDW transition at 207 K, the \la\ NMR spectra undergo sharp changes, establishing that a CDW gap opens up at the Fermi surface and two inequivalent \la\ sites are generated as a result of the periodic lattice distortion. These provide microscopic evidence that \LAS\ undergoes a Peierls transition, which is quite rare in non-1D materials. 

Our NMR studies strongly suggest that the square-net not only induces the band inversions being a good structural motif for topological semimetals that host 2D Dirac fermions, it could be also an excellent platform to study the CDW phenomenon in quasi-two dimensions and its relationship with Dirac fermions.

%


\begin{acknowledgments}
We thank Changyong Song for useful discussions. 
This work was supported by a National Research Foundation of Korea (NRF) grant funded by the Korea Government(MSIT) (Grant No.\,NRF-2020R1A2C1003817).
Work at the Ames Laboratory was supported by the U.S. Department of Energy, Office of Science, Basic Energy Sciences,
Materials Sciences and Engineering Division. The Ames Laboratory is operated for the U.S. Department of Energy by Iowa
State University under Contract No.\,DE-AC02-07CH11358. 
\end{acknowledgments}


\bibliography{mybib}

\end{document}